\documentclass[english,aps,prl,twocolumn]{revtex4-1}
\usepackage[T1]{fontenc}
\usepackage[latin9]{inputenc}
\usepackage{textcomp}
\usepackage{amsmath}
\usepackage{amssymb}
\usepackage{graphicx}
\usepackage{esint}

\makeatletter

\newcommand{\lyxmathsym}[1]{\ifmmode\begingroup\def\b@ld{bold}
  \text{\ifx\math@version\b@ld\bfseries\fi#1}\endgroup\else#1\fi}

\@ifundefined{textcolor}{}
{%
 \definecolor{BLACK}{gray}{0}
 \definecolor{WHITE}{gray}{1}
 \definecolor{RED}{rgb}{1,0,0}
 \definecolor{GREEN}{rgb}{0,1,0}
 \definecolor{BLUE}{rgb}{0,0,1}
 \definecolor{CYAN}{cmyk}{1,0,0,0}
 \definecolor{MAGENTA}{cmyk}{0,1,0,0}
 \definecolor{YELLOW}{cmyk}{0,0,1,0}
 }

\usepackage{hyperref}

\makeatother

\usepackage{babel}
\begin{document}

\title{Non-Markovian Dynamical Maps:\\Numerical Processing of Open Quantum
Trajectories }

\author{Javier Cerrillo and Jianshu Cao}

\email{jianshu@mit.edu}

\selectlanguage{english}%

\address{Massachusetts Institute of Technology, 77 Massachusetts Avenue, Cambridge
MA 02139.}
\begin{abstract}
The initial stages of the evolution of an open quantum system encode
the key information of its underlying dynamical correlations, which
in turn can predict the trajectory at later stages. We propose a general
approach based on non-Markovian dynamical maps to extract this information
from the initial trajectories and compress it into non-Markovian transfer
tensors. Assuming time-translational invariance, the tensors can be
used to accurately and efficiently propagate the state of the system
to arbitrarily long time scales. The non-Markovian transfer tensor
method (TTM) demonstrates the coherent-to-incoherent transition as
a function of the strength of quantum dissipation and predicts the
non-canonical equilibrium distribution due to the system-bath entanglement.
TTM is equivalent to solving the Nakajima-Zwanzig equation, and therefore
can be used to reconstruct the dynamical operators (the system Hamiltonian
and memory kernel) from quantum trajectories obtained in simulations
or experiments. The concept underlying the approach can be generalized
to physical observables with the goal of learning and manipulating
the trajectories of an open quantum system.
\end{abstract}
\maketitle
\textit{Introduction.---}The dynamics of large open quantum systems
are of interest to a broad range of disciplines, including condensed
matter physics, ultrafast spectroscopy, and quantum information technology,
just to name a few. Of particular interest is the interaction between
the system under study and the environment to which it couples. Within the fast bath approximation, the evolution of the system's density matrix is dictated
by a Lindbladian superoperator and can be regarded as a linear Markovian
process. Nevertheless, in general
the quantum trajectory of the open system is entangled with the bath
and is therefore temporally correlated, i.e., non-Markovian. The analysis
and simulation of this correlation is a daunting task, which often
requires resources that scale exponentially with the system size.
The root of the problem is the lack of a compact but complete representation
of the information encoded in open quantum trajectories. The standard
approaches fall into two classes: quantum master equations and path
integral simulations. The first class of approaches is based on formally
exact equations of motion, such as the Nakajima-Zwanzig formalism
\cite*{Breuer2007,Nakajima1958,Zwanzig1960} or others, but restricted to either weak damping, high-temperature, short memory time or short simulation time \cite{Ishizaki2005,Cao1997,Koch2003,Hughes2009,Gualdi2013,Breuer2004,Piilo2008}. The second class
of approaches adopts the harmonic bath assumption which renders the
use of stochastic Gaussian sampling or influence functional possible
\cite{Egger2000,Makri1995}, but does not converge well with the
system size, the length of the memory time or the strength of the
dissipation. To overcome these difficulties, we need a radically different
approach to dissipative quantum dynamics.

In this letter, we propose a unified method to characterize, reconstruct,
and propagate quantum trajectories which are non-locally correlated
in time. It applies to any form of the system-bath Hamiltonian and
scales favorably with respect to the system size and length of the
time correlation. The scheme is based on a black-box analysis that
extracts all available information from samples of initial trajectories
generated experimentally or numerically. This information is stored
in a collection of non-Markovian dynamical maps, which describes the
propagation of the initial state of the system to a later time with
full account for the time correlations in the trajectories. Then,
a transformation of these maps is performed to obtain a set of transfer
tensors that sort out correlations over different times. These tensors
are the central object of this letter and serve two purposes: On the
one hand, under the assumption of time-translational invariance, the
tensor formalism can be used in a multiplicative fashion to propagate
the state of the system to arbitrarily long times, leading to a non-perturbative
and efficient algorithm for simulating dissipative quantum systems.
On the other hand, the tensor multiplication method can be identified
as the formal solution to the Nakajima-Zwanzig equation, such that
one can reconstruct the system Hamiltonian and the memory kernel from
the tensors, and therefore design a procedure for non-Markovian quantum
process tomography.

\textit{Extraction of non-Markovian dynamical maps}.---The concept of dynamical maps \cite{Choi1975} has been extensively
explored as it contains all possible information on a quantum dynamical
system \cite{Chruscinski2011,Rivas2010,Breuer2009}. It is known from studies in quantum process tomography
\cite{Nielsen2000,Altepeter2003,Mohseni2008} that it is possible
to obtain the dynamical map  by adopting the concept of black-box engineering. 
The standard approach is to initialize the dynamics with a complete
basis set of the Hilbert space and then perform an input-output analysis
of the propagation. Here we apply this approach to non-Markovian
open quantum trajectories to generate the dynamical maps  at the discretized times
$t_{k}=k\cdot\delta t$, where $\delta t$ is the time step of the
simulation or the time resolution of the experiment,
\begin{equation}
\rho(t_{k})=\mathcal{E}_{k}\rho(0).\label{eq:QPT}
\end{equation}
The initial condition of the map is the identity operator, $\mathcal{E}_{0}=I$. Below we show
that, by decoding the information contained in the finite set of dynamical
maps $\left\{ \mathcal{E}_{k}\right\} $, one can develop an efficient
method to learn, propagate, and reconstruct the dynamics of an open
quantum system.

\begin{figure}
\begin{centering}
\includegraphics[bb=50bp 0bp 842bp 595bp,clip,height=0.25\textheight]{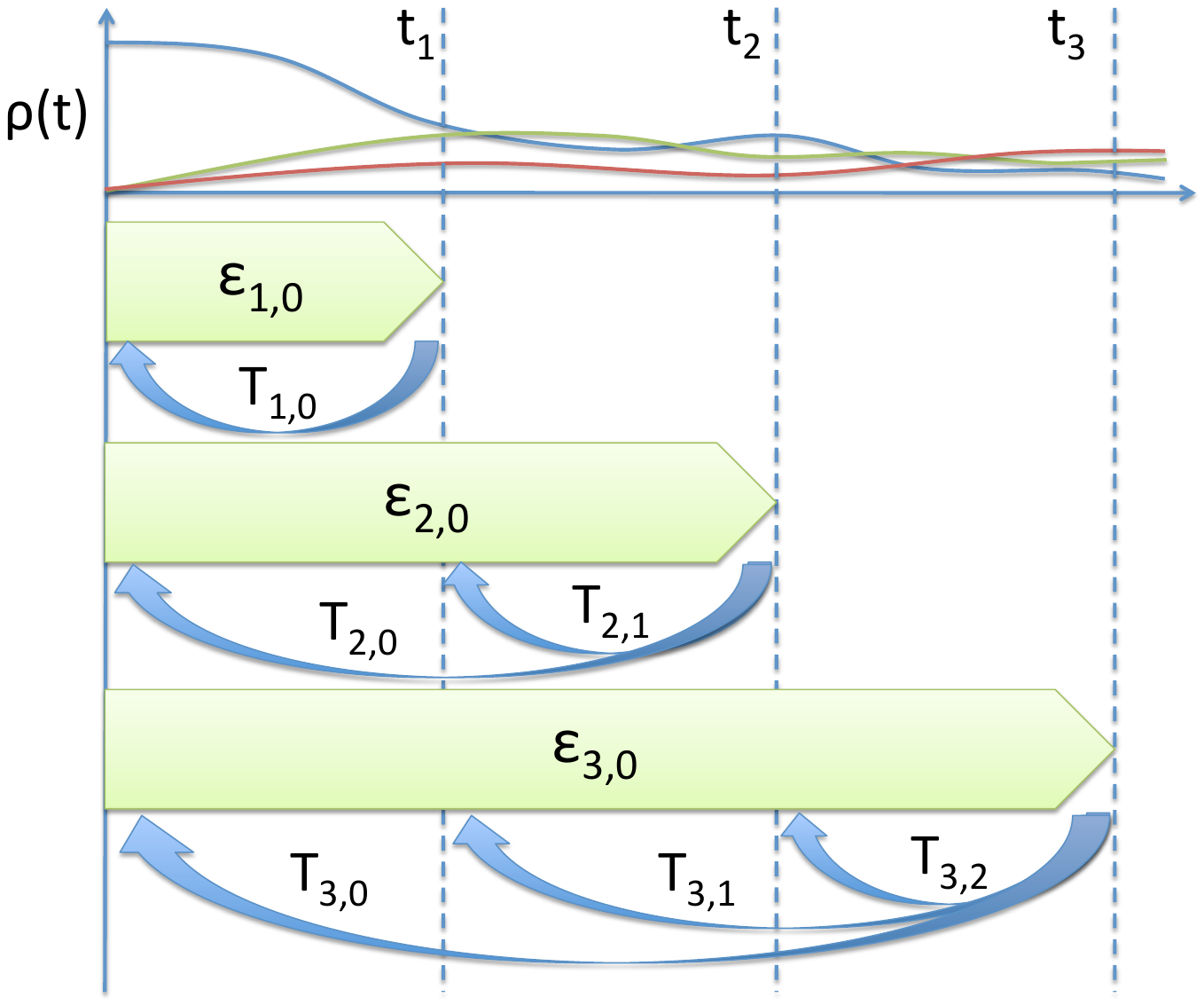}
\par\end{centering}

\caption{\label{fig:T-E}Diagram of the relationship between the dynamical
maps $\mathcal{E}_{k}$ and the transfer tensors $\mathrm{T}_{k,r}$.
While the dynamical map propagates any initial condition to a later
time $t_{k}$, the transfer tensor $\mathrm{T}_{k,r}$ quantifies
the direct correlation of the state of the system at time $t_{k}$
and the state at time $t_{r}$.}
\end{figure}

\textit{Propagation via tensor multiplication}.---Among different possible definitions \cite{Rivas2010,Breuer2009}, in this paper we identify non-Markovianity with violation of the semi group property. If the evolution
is Markovian it is possible to use the same map to propagate over
longer times in a multiplicative fashion, i.e. $\mathcal{E}_{n}=\mathcal{E}_{1}^{n}$
. Examples include conserved quantum dynamics and time-local dissipative
master equations (i.e. those of Lindblad form \cite{Lindblad1975}).
In a non-Markovian process each dynamical map needs to be found independently,
which constitutes a highly inefficient task since it contains correlations of the state of
the system at the present time with the states at all the previous
time steps. Here we transform the set of dynamical maps into a set
of transfer tensors $\mathrm{T}$ such that, regardless of the degree
of Markovianity of the environment, one can always propagate the system
in a multiplicative fashion. For this we propose the following transformation
\begin{equation}
\mathrm{T}_{n,0}=\mathcal{E}_{n}-\sum_{m=1}^{n-1}\mathrm{T}_{n,m}\mathcal{E}_{m},\label{eq:Ts}
\end{equation}
which reduces to the Markovian limit if all $\mathrm{T}$ vanish except
for the tensors corresponding to a single timestep $\mathrm{T}_{n+1,n}$. Illustrated in Fig.(\ref{fig:T-E}),
this expression establishes the relationship between the maps $\mathcal{E}_{n}$
and the tensors $\mathrm{T}_{n,m}$, and allows us to transform the
dynamical mapping defined in Eq.(\ref{eq:QPT}) into a dynamical propagation,

\begin{equation}
\rho(t_{n})=\sum_{k=0}^{n-1}\mathrm{T}_{n,k}\rho(t_{k}).\label{eq:Iterative-1}
\end{equation}
In a sense, we regard the non-Markovian effect as time correlations
in the quantum trajectory, and encode the correlation between any
pair of time slices $t_{k}<t_{n}$ in the tensor $\mathrm{T}_{n,k}$,
so that $\mathrm{T}_{n,k}\rho\left(t_{k}\right)$ corresponds to the
component of $\rho\left(t_{n}\right)$ that is conditioned on $\rho\left(t_{k}\right)$.
As in Eq.(\ref{eq:QPT}), the summation over all the possible components
determines the density matrix at time $t_{n}$.

The formulation presented above is general, and will now be simplified
by assuming time-translational invariance and finite time correlation
in the transfer tensor. Under certain assumptions, e.g. separable
system-bath initial conditions and a time independent Hamiltonian, we may invoke time-translational invariance
so that the transfer tensor is a function of the time difference only,
$\mathrm{T}_{n,k}=\mathrm{T}_{n-k}$. This selects a unique time frame in the dynamics, related for instance
to the initial pulse in ultrafast laser excitation. We note that the
time-translational property applies in general only to the transfer tensor, whereas
for the dynamical map it holds only under Markovian or unitary dynamics. The tensors may now be obtained from the set of dynamical maps in an
iterative fashion. $\mathcal{E}_{1}$ corresponds
to the Markovian part of the dynamical map, generated by the system
Hamiltonian and the initial relaxation due to the system-bath interaction,
and we define $\mathrm{T}_{1}\equiv\mathcal{E}_{1}$. The second map
$\mathcal{E}_{2}$ contains the effect of two sequential Markovian
steps, $\mathcal{E}_{1}^{2}$. Any deviation of $\mathcal{E}_{2}$
from $\mathcal{E}_{1}^{2}$ arises from the memory effect of the non-Markovian
bath and is encapsulated in the transfer tensor $\mathrm{T}_{2}\equiv\mathcal{E}_{2}\lyxmathsym{\textendash}\mathcal{E}_{1}\mathcal{E}_{1}=\mathcal{E}_{2}-\mathrm{T}_{1}\mathcal{E}_{1}$.
Similarly, $\mathcal{E}_{3}$ contains the correlation between $t_0$ and $t_{3}$, and $\mathrm{T}_{3}\equiv\mathcal{E}_{3}\lyxmathsym{\textendash}\mathrm{T}_{1}\mathcal{E}_{2}\lyxmathsym{\textendash}\mathrm{T}_{2}\mathcal{E}_{1}$.
Following this procedure, one can iteratively extract past-present
correlations from dynamical maps. Since the timespan of bath correlations
in realistic systems is finite, one may define a
cutoff $K$ such that $\mathrm{T}_{s}\rightarrow0$ for $s>K$.  In practice, one would define an accuracy threshold $\epsilon$ for some measure of the magnitude of the tensor (for instance, the trace norm). One would define the cutoff as the point where that measure falls below the threshold. This
justifies the truncation of the sum in (\ref{eq:Iterative-1}) at
$k=K$. Therefore, the dynamics of a large class of quantum systems
may be encoded in the finite set of transfer tensors $\mathrm{T}_{s}$
with $s\in\left\{ 1,\dots,K\right\} $  and the matrix propagation equation (\ref{eq:Iterative-1}) for
times $t_{n}>t_{K}$ can be regarded as a tensor multiplication method,
\[
\rho(t_{m})=\left(\begin{array}{cccc}
\mathrm{T}_{1} & \mathrm{T}_{2} & \ldots & \mathrm{T}_{K}\end{array}\right)\left(\begin{array}{c}
\rho(t_{m-1})\\
\rho(t_{m-2})\\
\vdots\\
\rho(t_{m-K})
\end{array}\right),
\]
which is the non-Markovian extension of time-local dissipative quantum
dynamics and defines our Transfer Tensor Method (TTM) for propagation.

The TTM equation completes the basic 4-steps scheme: generate short-time
trajectories numerically or experimentally, learn from short-time
trajectories to extract the dynamical maps in Eq.(\ref{eq:QPT}),
use Eq.(\ref{eq:Ts}) to derive the transfer tensors from the map,
and evolve the density matrix to arbitrarily long time according to
Eq.(\ref{eq:Iterative-1}). This procedure is completely general,
and applies to any system or bath, continuous or discrete. Yet, for
simplicity of the benchmarking, we will use the numerical example
of the spin-boson model below.

\textit{Scaling and error estimation}.--- With TTM we hope to address a challenge faced by many numerically exact simulation methods of open quantum systems: one can typically recognize an unfavorable exponential scaling of resources with the number of
simulated time-steps. In the case of stochastic or Monte Carlo methods, an exponentially larger sample size is required for longer simulations. Renormalization methods \cite{Koch2003,Hughes2009,Gualdi2013} require an ever increasing representation of the bath as time increases. The hierarchy of equations of motion \cite{Ishizaki2005} is limited to the
Drude-Lorentz bath and is characterized by a factorial scaling with the
increase of the hierarchy levels. In the case
of QUAPI \cite{Makri1995} or any other deterministic approach based
on a path integral formulation, each path requires explicit
storage and the tensor scales exponentially like $D^{2(K+1)}$ for a Hilbert space of dimension $D$ and a truncation
$K$ on the memory kernel. The TEDOPA \cite{Prior2010} method
has a more favorable scaling, but the size of its representation increases
with temperature and time. In all the cases mentioned, this time-dependent scaling is usually in addition to the one depending on the system size.
A remarkable aspect of the tensor multiplication method is the linear scaling of the storage
requirements, since the total size of the set of transfer tensors is
$KD^{4}$. Thus, by combining our method with an exact simulation one manages to reduce the required resources for long-time simulation by a significant amount. Based on this, our method is especially
suitable for large systems, strong damping, and long memory time. Another useful aspect of the scheme is the direct relationship between
the magnitude of the elements of the
last tensor and the accuracy of the propagation. Despite being system specific, a reasonable upper bound
for the error $e$ associated with the truncation at level $K$ is
represented by some norm measure of the first neglected correlation
operator, i.e. $e\simeq|T_{K+1}|$. This error accumulates in a multiplicative
fashion. Since $K$ can be arbitrarily chosen, the error of the long
time prediction can be reduced at will. It is worth noting that this
method is deterministic and is therefore not affected by the so-called
{}``sign problem'' of Monte Carlo and stochastic propagation methods.

\begin{figure}
\begin{centering}
\includegraphics[width=1\columnwidth]{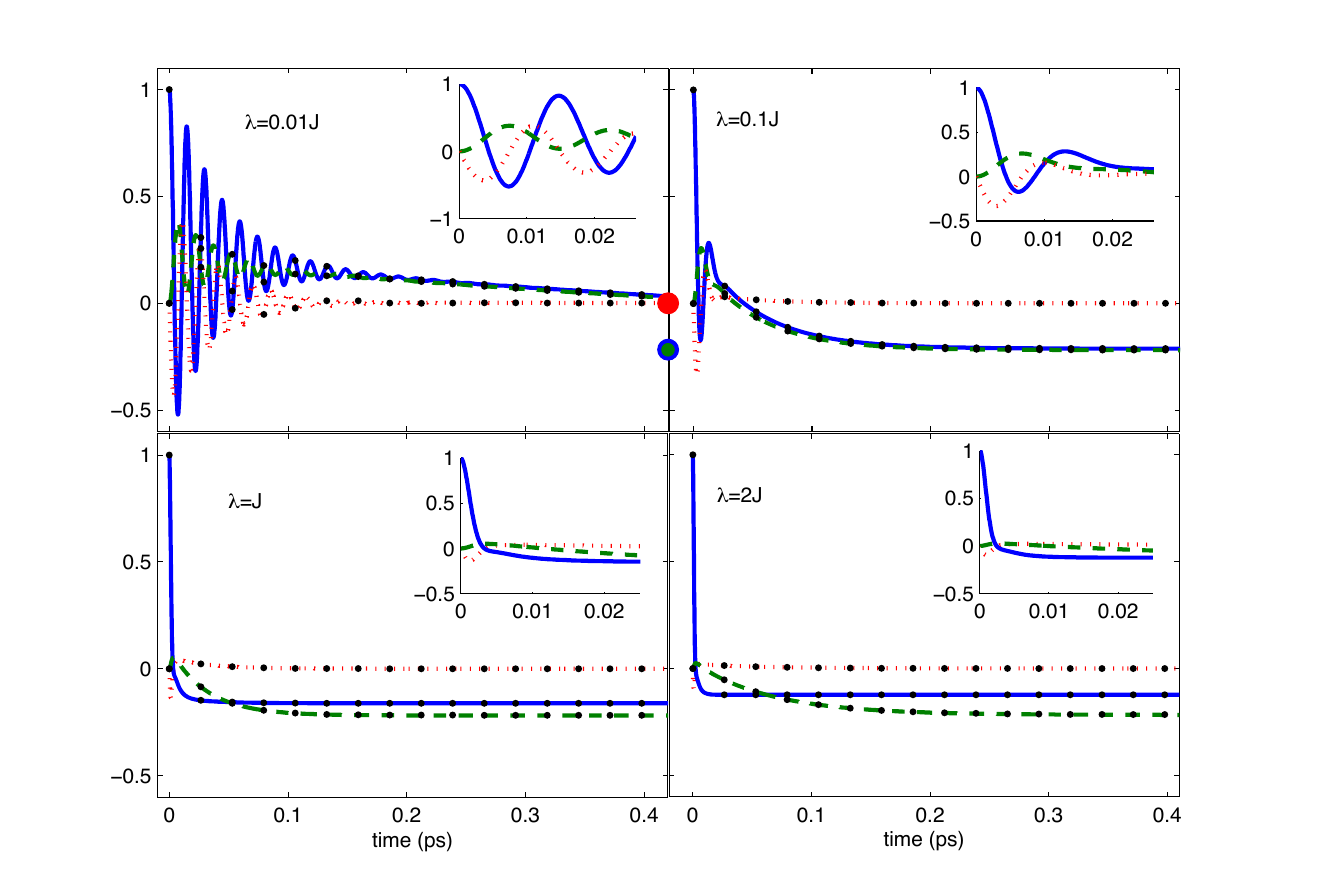}
\par\end{centering}

\caption{\label{fig:t1}Numerically exact simulation with the hierarchic method
(lines) and prediction with TTM (dots) of the density matrix elements
of a two level system under the influence of a Drude-Lorentz bath
for different values of the system-bath coupling $\lambda$. Blue
(solid) lines correspond to $\left(\rho_{11}-\rho_{22}\right)$, green
(dotted) lines to $Re\left\{ \rho_{12}\right\} $ and red (dashed)
lines to $Im\left\{ \rho_{12}\right\} $. The inset of every plot
shows the corresponding learning period for the initial condition
$\rho_{11}(0)=1$. The system-bath parameters are $T=300K$ ($k_{B}T\simeq2\hbar J$),
and $\gamma=J$, with $J=100cm^{-1}$. The equilibrium state for the
case of $\lambda=0.01J$ is shown with colored dots on the axis.}
\end{figure}

\begin{figure}
\begin{centering}
\includegraphics[width=.8\columnwidth]{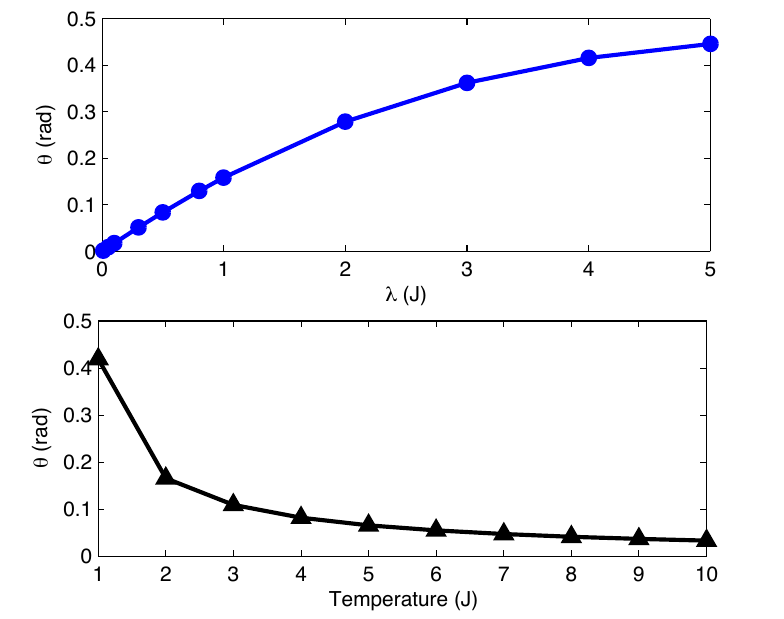}
\par\end{centering}

\caption{\label{fig:t2}Deviation $\theta$ from the canonical distribution
as a function of the system-bath coupling $\lambda$, the cutoff frequency
$\gamma$ and the temperature $T$. The learning period spans the
first 25fs of the dynamics. The coupling is $J=100cm^{-1}$ and the
cutoff frequency is $\gamma=5J$. In the upper panel $k_{b}T=2\hbar J$
and in the lower panel $\lambda=J$.}
\end{figure}

\textit{Examples of propagation}.---We now demonstrate the applicability
of the proposed method with two examples. To begin with, trajectories
of a biased two level system (TLS) with exponentially decaying noise
have been generated using the hierarchy method \cite{Ishizaki2005}
and the TTM. The frequency difference of the TLS is $\omega_{0}=100\mathrm{cm^{-1}}$
and the intrinsic coupling is $J=\omega_{0}$. The TLS couples off-diagonally
to a harmonic bath of temperature $T=300K$, a characteristic frequency
of $\gamma=J$ and a variety of system-bath couplings ranging from
$\lambda=0.01J$ to $\lambda=2J$. As shown in figure (\ref{fig:t1}),
after the initial learning period shown in the insets, TTM successfully
reproduces the transient dynamics of the density matrix until it reaches
equilibrium. The different values of system-bath couplings are chosen
such that the crossover between underdamped to overdamped dynamics
is illustrated: while the first panel with $\lambda=0.01J$ contains
oscillatory dynamics at short time, these progressively disappear
with increasing $\lambda$ until no trace of coherent oscillations
can be observed for $\lambda=2J$. It is worth noting that the composition
of the equilibrium state varies with $\lambda$ due to quantum noncanonical
statistics \cite{Lee2012,Moix2012}. This effect is correctly reproduced by
our method, which indicates its suitability to predict long time dynamical
features with high accuracy. As the second example, we have further
explored this effect quantitatively using a measure of the deviation
from the canonical equilibrium distribution as defined in \cite{Lee2012}:
the Bloch sphere angular distance $\theta$ between the canonical
distribution eigenbasis and the non-canonical one. Figure (\ref{fig:t2})
explores the deviation $\theta$ as a function of the system-bath
coupling $\lambda$ and the temperature $T$. The equilibrium entanglement
with the bath increases with the strength of the interaction, which
results in an increasing deviation from the canonical equilibrium
that is specific to quantum systems. For a strong enough system-bath
coupling, the non-canonical distribution reaches the eigenbasis of
the system-bath interaction operator and the deviation angle saturates.
In the present case, the saturation angle corresponds to $\frac{\pi}{4}$
radians. In contrast, thermalization supresses entanglement, which
explains the decrease of deviation from the canonical equilibrium
with increasing temperature. The equilibrium density matrix returns
to the Boltzmann distribution in the high-temperature limit. It is
evident from these two examples that TTM is suitable for long time
simulation of dissipative quantum systems for which existing methods
are not efficient or practical. This opens up a new possibility for
exploring a plethora of previously unaccessible dynamical regimes.

\textit{Extraction of the Nakajima-Zwanzig equation}.---A unique aspect
of the TTM approach is the possibility of using the complete information
obtained via the dynamical maps to generate the equation of motion
and the associated dynamical operators. Earlier, simple methods have
been used to determine the Hamiltonian and the Markovian decoherence
\cite{Cina2006}. Here we demonstrate that the non-Markovian
nature of the dynamics, i.e. the memory kernel and consequently the
correlation function of the environment, can also be rigorously determined.
For this reason, it is useful to relate the non-Markovian dynamical
maps to the physical terms in the exact Nakajima-Zwanzig formalism
\cite{Breuer2007,Nakajima1958,Zwanzig1960}, where the evolution of
an open quantum system with separable initial conditions can be expressed
exactly in the form of the equation

\begin{equation}
\dot{\rho}(t)=-i\mathcal{L}_{s}\rho(t)+\int_{0}^{t}\mathcal{K}(t,t')\rho(t')dt'.\label{eq:Integrodifferential}
\end{equation}
Here, $\mathcal{L}{}_{s}$ is the Liouvillian of the system alone,
and $\mathcal{K}(t,t')$ is the memory kernel due to the system-bath
interaction \citep{Breuer2007}. This equation can be seen as the continuous limit of Eq. (\ref{eq:Iterative-1}).
Thus, by comparing the time\textendash{}convoluted kernels, we can
easily identify,
\begin{equation}
\mathrm{T}_{k,n} =(1-i\mathcal{L}_{s}\delta t)\delta_{k,n+1}+\mathcal{K}_{k,n}\delta t^{2}.
\end{equation}
Here $\mathcal K _{a,b}=\mathcal K (t_a,t_b)$ and $\delta_{a,b}$ is the Kronecker delta. This identity not only elucidates the physical motivation
for using the transfer tensors $\mathrm{T}_{k}$ instead of the dynamical
maps $\mathcal{E}_{k}$ in the numerical propagation of the density
matrix, but also suggests using TTM to evaluate the memory kernel.
Specifically, we extract the dynamical maps from the short-time dynamics,
use Eq.(\ref{eq:Ts}) to transform the maps into tensors, and then
identify the tensors with the system Hamiltonian and memory kernel
in Eq.(\ref{eq:Integrodifferential}).

\begin{figure}
\begin{centering}
\includegraphics[width=.8\columnwidth]{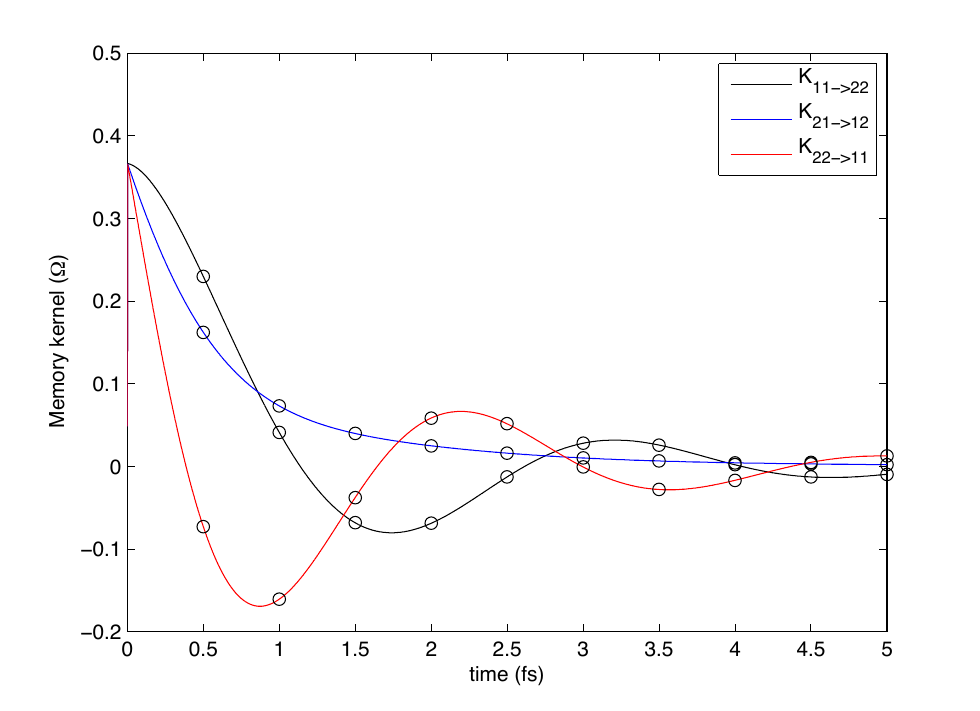}
\par\end{centering}

\caption{\label{fig:correlation}Real parts of non-zero elements of the memory
kernel as a function of time. The nonzero elements are $\mathcal{K}_{11\rightarrow11}=-\mathcal{K}_{11\rightarrow22}$,
$\mathcal{K}_{22\rightarrow22}=-\mathcal{K}_{22\rightarrow11}$ and
$\mathcal{K}_{12\rightarrow21}=-\mathcal{K}_{21\rightarrow21}=-\mathcal{K}_{12\rightarrow12}=\mathcal{K}_{21\rightarrow12}$.
The solid lines represent the matrix elements of the memory Kernel
extracted by applying TTM to a hierarchy simulation of a symmetric
two level system with coupling $\Omega$. The dots are the values
corresponding to the analytical prediction. The dissipative system
is characterized by the parameters $\lambda=0.25\Omega$, $\gamma=0.05\Omega$
and $\beta\Omega=4.79$.}
\end{figure}

An example that illustrates this approach is shown in Fig.(\ref{fig:correlation}),
where the memory kernel is plotted as a function of time for the spin-boson
model. The numerical results are extracted from a hierarchy simulation
of short-time trajectories under the influence of a Drude-Lorentz
bath. As the benchmark, we compare the simulated memory kernel with
the prediction from path integral calculations using the Feynman-Vernon
influence functional and find perfect agreement. The ability to generate
the transfer tensors $\mathrm{T}_{k}$ directly from the influence
functional for large-scale simulations of quantum dissipative systems
(up to hundreds of excitonic states) will be presented elsewhere.

The connection to the memory kernel further specifies the conditions under which time-translational invariance can be assumed. If (i) the total Hamiltonian is time-independent, (ii) the initial total state is a product state and (iii) the initial state of the environment is a stationary state, then the memory kernel depends only on the difference of its arguments, i.e. $\mathcal{K}(t,t')=\mathcal{K}(t-t',0)$; see \S 5.1 in \cite{Rivas2012}, also \cite{Breuer2007,Cao1997}. These conditions are hence directly related to the time-translational invariance of the transfer tensors.

\textit{Conclusion}.---We have presented a strategy based on non-Markovian
dynamical maps to process the relevant information encapsulated in
the trajectory of an open quantum system. This information can be
used to learn about the underlying dynamics of the system in order
to generate a set of transfer tensors for propagation to longer time
scales. Applications of TTM to short-time trajectories clearly demonstrate
the dynamic transition from coherent oscillations to incoherent transfer
and accurately predict the non-canonical equilibrium distributions.
Further, the transfer tensor method can be used to reconstruct the
relevant dynamical operators of the system such as its Hamiltonian
and memory kernel, by identifying the tensors with the Nakajima-Zwanzig
equation. Due to its adaptability and scalability, the proposal not
only constitutes a pedagogic approach to the description of open quantum
systems but also stands out as a promising technique to extend the
timespan of simulating quantum trajectories. 
\begin{acknowledgments}
This work was supported by grants from the National Science Foundation
(Grant No. CHE-1112825), DARPA, and DOE. Javier Cerrillo is currently
supported by the Center for Excitonics at MIT funded by the Department
of Energy (Grant No. DE-SC0001088). 
\end{acknowledgments}
\bibliographystyle{apsrev4-1}
\bibliography{NPP2}

\end{document}